\RequirePackage{fixltx2e}
\documentclass[aps,jcp,reprint,floatfix]{revtex4-1}
\usepackage{amsmath,amsfonts,amssymb} %
\usepackage{mathtools} %
\usepackage{wasysym} %
\usepackage{bm} %
\usepackage[bbgreekl]{mathbbol} %
\usepackage{graphicx} %
\usepackage[dvipsnames,table]{xcolor} %
\usepackage{tabularx} %
\usepackage[acronym]{glossaries} %
\usepackage{braket} %
\usepackage{multirow} 

\usepackage{fancyhdr} %
\usepackage{microtype} %
\usepackage{natmove}

\usepackage[byname]{smartref} %
\usepackage[breaklinks, %
colorlinks, linkcolor=Blue, citecolor=Blue, urlcolor=Blue]{hyperref} %
\usepackage[version=3]{mhchem}
\AtBeginDocument{\usepackage{booktabs}} 

\newcommand{\RR}{\mathbf{R}} %
\newcommand{\BC}{\color{black} } %
\newcommand{\eq}[1]{Eq.~(\ref{#1})} %
\newcommand{\fig}[1]{Fig.~\ref{#1}} %

\def\be{\begin{equation}} %
\def\ee{\end{equation}} %
\def\bea{\begin{eqnarray}} %
\def\eea{\end{eqnarray}} %

\newacronym{QPE}{QPE}{quantum phase estimation} %
\newacronym{VQE}{VQE}{variational quantum eigensolver} %
\newacronym{UCC}{UCC}{unitary coupled cluster} %
\newacronym{QCC}{QCC}{qubit coupled cluster} %
\newacronym{FCI}{FCI}{full configurational interaction} %
\newacronym{CASCI}{CASCI}{complete active space configurational
  interaction} %
\newacronym{JW}{JW}{Jordan--Wigner} %
\newacronym{BK}{BK}{Bravyi--Kitaev} %
\newacronym[longplural={degrees of freedom}, %
firstplural={degrees of freedom (DOF)}, plural={DOF}]{DOF}{DOF}{degree
  of freedom} %
\newacronym[longplural={equations of motion}, %
firstplural={equations of motion (EOM)}, %
plural={EOM}]{EOM}{EOM}{equation of motion} %
\newacronym{PES}{PES}{potential energy surface} %
\newacronym{CI}{CI}{configuration interaction} %
\newacronym{QMF}{QMF}{qubit mean-field} %
\newacronym{SQP}{SQP}{sequential quadratic programming} %
\newacronym{RHF}{RHF}{restricted Hartree--Fock}

\begin{document}

\author{Vladyslav Verteletskyi${}^{a,b,c}$} 
\affiliation{${}^a$ Department of Physical and Environmental Sciences,
  University of Toronto Scarborough, Toronto, Ontario, M1C 1A4,
  Canada; ${}^b$Chemical Physics Theory Group, Department of Chemistry,
  University of Toronto, Toronto, Ontario, M5S 3H6, Canada; ${}^c$Department of Quantum Field Theory, Taras Shevchenko National University of Kyiv, Kyiv, 03022, Ukraine}

\author{Tzu-Ching Yen${}^b$} 
\affiliation{${}^a$ Department of Physical and Environmental Sciences,
  University of Toronto Scarborough, Toronto, Ontario, M1C 1A4,
  Canada; ${}^b$Chemical Physics Theory Group, Department of Chemistry,
  University of Toronto, Toronto, Ontario, M5S 3H6, Canada; ${}^c$Department of Quantum Field Theory, Taras Shevchenko National University of Kyiv, Kyiv, 03022, Ukraine}
  
\author{Artur F. Izmaylov${}^{a,b}$} 
\email{artur.izmaylov@utoronto.ca}
\affiliation{${}^a$ Department of Physical and Environmental Sciences,
  University of Toronto Scarborough, Toronto, Ontario, M1C 1A4,
  Canada; ${}^b$Chemical Physics Theory Group, Department of Chemistry,
  University of Toronto, Toronto, Ontario, M5S 3H6, Canada; ${}^c$Department of Quantum Field Theory, Taras Shevchenko National University of Kyiv, Kyiv, 03022, Ukraine}  
  
\title{Measurement Optimization in the Variational Quantum Eigensolver Using a Minimum Clique Cover}
\date{\today}

\begin{abstract}
Solving the electronic structure problem using the Variational Quantum Eigensolver (VQE) 
technique involves measurement of the Hamiltonian expectation value. Current hardware can perform 
only projective single-qubit measurements, and thus, the Hamiltonian expectation value 
is obtained by measuring parts of the Hamiltonian rather than the full Hamiltonian. This restriction 
makes the measurement process inefficient because the number of terms in the Hamiltonian  
grows as $O(N^4)$ with the size of the system, $N$. To optimize VQE measurement one can try to 
group as many Hamiltonian terms as possible for their simultaneous measurement. Single-qubit measurements
allow one to group only the terms commuting within corresponding single-qubit 
subspaces or qubit-wise commuting. We found that qubit-wise commutativity between the Hamiltonian terms 
can be expressed as a graph and the problem of the optimal grouping is equivalent of finding a 
minimum clique cover (MCC) for the Hamiltonian graph. The MCC problem is NP-hard but there exist several 
polynomial heuristic algorithms to solve it approximately. Several of these heuristics 
were tested in this work for a set of molecular electronic Hamiltonians. 
On average, grouping qubit-wise commuting terms reduced the number of operators to measure three times 
compared to the total number of terms in the considered Hamiltonians.  
\end{abstract}

\glsresetall

\maketitle

\section{Introduction}

The \gls{VQE} method\cite{Yung:2014iv,Peruzzo:2014/ncomm/4213, Jarrod:2016/njp/023023, Wecker:2015/pra/042303,Olson:2017ud,McArdle:2018we} is currently the most practical scheme for solving the electronic structure problem 
on current and near-future universal quantum computers. This method involves iterative optimization 
of the electronic energy 
\bea
E_e(\RR) = \min_{\ket{\Psi(\RR)}} \bra{\Psi(\RR)} \hat H_e (\RR) \ket{\Psi(\RR)}
\eea
using both quantum and classical 
computers. Here, $\RR$ is the studied nuclear configuration, $\hat H_e(\RR)$ is the electronic Hamiltonian,
and $\ket{\Psi(\RR)}$ is the electronic wavefunction. In \gls{VQE}, the quantum computer obtains 
Hamiltonian expectation values for trial wavefunctions suggested by the classical computer,
while the minimization process is done on the classical computer.  

At a more detailed level, the quantum computer (QC) does not work with $\hat H_e(\RR)$ but rather with 
its qubit counterpart ($\hat H_q$) obtained from the second quantized version of $\hat H_e(\RR)$.
\cite{Jordan:1928/zphys/631,Bravyi:2002/aph/210, Seeley:2012/jcp/224109,
Tranter:2015/ijqc/1431, Setia:2017/ArXiv/1712.00446,Havlicek:2017/pra/032332}  
To obtain the expectation values for $\hat H_q$, QC creates 
a quantum state of an artificial qubit system $\ket{\Psi_q}$ that emulates $\ket{\Psi(\RR)}$ 
and performs projective measurements on $\ket{\Psi_q}$. 
A typical system qubit Hamiltonian has the form       
\begin{equation}
  \label{eq:spin_ham}
  \hat H_q = \sum_I C_I\,\hat P_I,
\end{equation}
where $C_I$ are numerical coefficients, and $\hat P_I$ are
Pauli  ``words", products of Pauli operators of different qubits 
\begin{equation}
  \label{eq:Pi}
  \hat P_I = \prod_{i=1}^{N} \hat \sigma_{i}^{(I)},
\end{equation}
$\hat \sigma_i^{(I)}$ is one of the $\hat x,\hat y,\hat z$ Pauli operators or identity $\hat e$ for the $i^{\rm th}$ qubit.
The number of qubits $N$ is equal to the number of spin-orbitals used in the second quantized form 
of $\hat H_e$, and the total number of Pauli words in $\hat H_q$ scales as $N^4$ due to the two-electron 
integral component of $\hat H_e$ in second quantization. 

Measuring the whole qubit Hamiltonian [\eq{eq:spin_ham}] is not currently technologically possible in this setup. 
This is quite different from the quantum simulator model, where the Hamiltonian of the system of interest 
is modelled by another tuneable quantum system which is amenable to eigen-spectrum 
measurements.\cite{cirac:2012,cirac:2018} 
Instead, within \gls{VQE}, only single-qubit operators $\hat \sigma_i$ can be measured. Due to 
projective nature of these measurements one can determine eigenvalues of operators that 
share the same tensor product eigen-basis. For example, for a two-qubit system, results of 
$\hat z_1$, $\hat z_2$, and $\hat z_1\hat z_2$ measurement 
can be obtained by measuring $\hat z_1$ and $\hat z_2$ because for all these operators 
product states 
$\{\ket{\uparrow\uparrow},\ket{\uparrow\downarrow},\ket{\downarrow\uparrow},\ket{\downarrow\downarrow}\}$
are eigenstates of single-qubit operators (e.g. $\hat z_1\ket{\uparrow\downarrow} = +1\ket{\uparrow\downarrow}, 
$ and $\hat z_2\ket{\uparrow\downarrow} = -1\ket{\uparrow\downarrow}$). However, $\hat x_1$, $\hat x_2$, and 
$\hat x_1\hat x_2$ would require a separate set of measurements. Interestingly, 
even though operators $\hat z_1\hat z_2$ and $\hat x_1\hat x_2$ commute and share 
the common system of eigenstates, they cannot be measured at the same time using 
single-qubit projective measurements. The problem is that their common eigenstates 
do not have a simple tensor product form in this case, instead they are entangled superpositions 
in both $\hat z_i$ or $\hat x_i$ single-qubit eigen-bases,
$\{(\ket{\uparrow\uparrow}\pm\ket{\downarrow\downarrow})/\sqrt{2}, 
(\ket{\uparrow\downarrow}\pm\ket{\downarrow\uparrow})/\sqrt{2}\}$.
  
 Most of $\sim N^4$ terms in $\hat H_q$ do not share 
one tensor product basis (TPB),\cite{Kandala:2017/nature/242}  
moreover, as we will show later there is no unique partitioning 
to groups of terms sharing TPB. This poses a question of how to minimize the number of groups 
whose terms can be measured simultaneously. 
Here, we address this problem by reformulating it as a minimum clique cover (MCC) problem for a 
graph representation of the system Hamiltonian. Previously, there were other attempts to 
address this problem either using variance estimates\cite{Jarrod:2016/njp/023023} or searching 
for optimal TPB sharing group partitioning by inspection.\cite{Kandala:2017/nature/242} 
However, it seems that their systematic application to Hamiltonians with thousands of terms can be 
problematic. Recently, graph based techniques similar to ours have been implemented in the 
Rigetti's pyQuil set of programs,\cite{Rigetti_doc} but no systematic description of their performance 
can be found in the literature. Thus, in this work we discuss the connection of the grouping 
problem with graph-based techniques and assess the performance 
of both Rigetti's algorithms and our developments on a set of qubit Hamiltonians for small 
molecules with up to 36 qubits and 53 thousands of terms.  

The rest of the paper is organized as follows. Section \ref{sec:measurability} provides the  
connection between the grouping of terms based on shared TPB and the MCC problem. Then,
we discuss multiple heuristic approaches to MCC in Sec.~\ref{sec:MCC}. Section~\ref{sec:numer-stud-disc} 
illustrates performance of the considered heuristic approaches on a set of qubit Hamiltonians. 
Section~\ref{sec:conclusions} concludes with a summary of main results.

\section{Theory}
\label{sec:theory}

\subsection{Simultaneously measurable fragments} 
\label{sec:measurability}

To formalize the condition of two terms $\hat P_I = \prod_i \hat \sigma_i$ 
and $\hat P_J = \prod_i \hat \sigma'_i$ sharing TPB, it is useful to introduce
qubit-wise commutativity as a zero value of qubit-wise commutator 
\bea
[\hat P_{I},\hat P_{J}]_{\rm qw} =\begin{cases}
      0, & \text{if}\ [\hat \sigma_i,\hat \sigma'_i] =0 ~\forall i \\
      1, & \text{otherwise} 
    \end{cases}.
\eea
Thus, $[\hat P_{I},\hat P_{J}]_{\rm qw}$ is zero only if all one-qubit operators in $\hat P_{I}$ 
commute with their counterparts in $\hat P_{J}$.
Clearly, if $\hat P_{I}$ and $\hat P_{J}$ qubit-wise commuting (QWC) then they commute in the normal sense 
$[\hat P_{I},\hat P_{J}]=0$. The 
opposite is not true, a simple example is $[\hat x_1\hat x_2, \hat y_1\hat y_2] = 0$ but 
$[\hat x_1\hat x_2, \hat y_1\hat y_2]_{\rm qw} \ne 0$.   

To partition the qubit Hamiltonian $\hat H_q$ into groups of terms sharing TPB, 
it is necessary and sufficient to group terms that mutually QWC, 
\bea\label{eq:split}
\hat H_q &=&  \sum_{n} \hat A_n, ~\hat A_n = \sum_{I} C_{I}^{(n)} \hat P_{I}^{(n)}, \\
&&{[}\hat P_{I}^{(n)},\hat P_{J}^{(n)}{]}_{\rm qw} = 0.
\eea
Partitioning of the $\hat H_q$ in Eq.~(\ref{eq:split}) allows one to measure all Pauli words within 
each $\hat A_n$ group in a single set of $N$ one-qubit measurements. For every qubit, it is known from the form of
$\hat A_n$, what Pauli operator needs to be measured. The advantage of this scheme is that 
it requires only single-qubit measurements, which are technically easier than multi-qubit measurements.
The disadvantage of this scheme is that the Hamiltonian may require to measure too many $\hat A_n$ 
terms separately.  A natural question arises: how to obtain the optimal grouping of terms to minimize the number 
of the $\hat A_n$ groups? 

This question is nontrivial because the QWC relation is not the equivalence relation in the 
algebraic sense and thus does not provide a unique non-overlapping partitioning to  
equivalence classes. To see this, let us recall that for the equivalence relation $\sim$
between elements of any set $\{a,b,c,...\}$ we need to have three 
conditions: 1) $a\sim a$, 2) if $a\sim b$ then $b\sim a$, and 3) if $a\sim b$ and $b\sim c$ then $a\sim c$. 
If these conditions are satisfied the set can be split into non-overlapping unique subsets 
whose elements are all equivalent to each other. Unfortunately, only conditions 1) and 2)       
are satisfied for the QWC relation, which is not enough for the equivalence relation. 
Indeed, violation of (3) by QWC is easy to see on the following example: $[\hat x_1,\hat y_2]_{\rm qw} = 0$ and 
$[\hat y_2,\hat z_1]_{\rm qw} = 0$ but that does not lead to $[\hat x_1,\hat z_1]_{\rm qw} = 0$.

\begin{figure}[h!]
  \centering %
  \includegraphics[width=0.7\columnwidth]{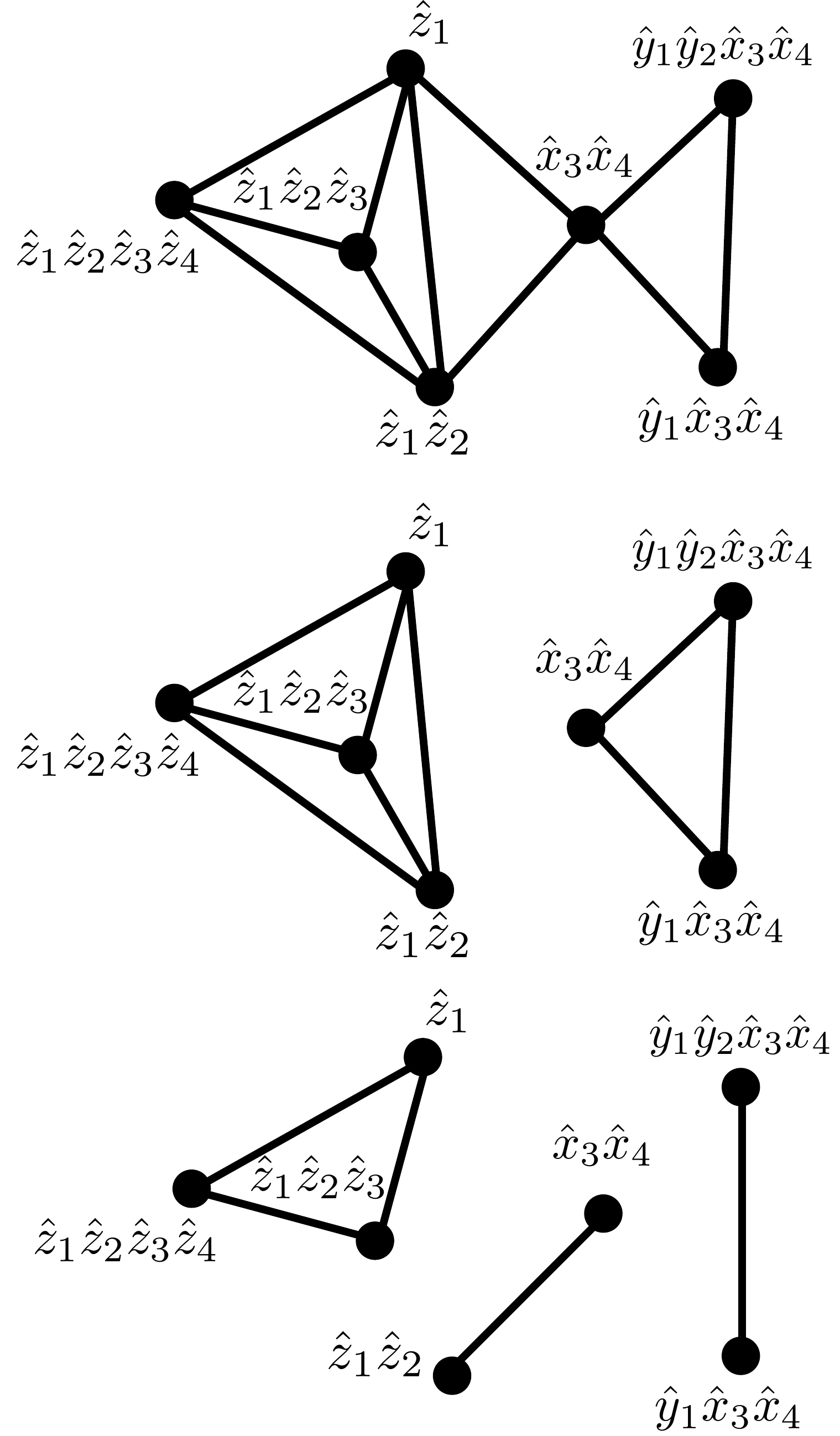}
  \caption{Graph representation of QWC terms in the Hamiltonian \eq{eq:Hex} (upper panel), minimum clique 
  cover of the graph (middle panel), non-minimum clique cover of the 
  graph (lower panel).}
  \label{fig:QWCG}
\end{figure}

Yet, the two conditions that are satisfied for the QWC relation are enough to represent 
the QWC relation as graph edges between the Pauli words of the Hamiltonian. As a simple illustration 
one can consider the following model Hamiltonian
\bea\notag
\hat H &=& \hat z_1 + \hat z_1\hat z_2 + \hat z_1\hat z_2\hat z_3 + \hat z_1\hat z_2\hat z_3\hat z_4 \\ \label{eq:Hex}
&&+ \hat x_3\hat x_4 + \hat y_1\hat x_3\hat x_4 + \hat y_1\hat y_2\hat x_3\hat x_4,
\eea
whose QWC graph is given in Fig.~\ref{fig:QWCG}. To determine how many terms can be measured 
simultaneously, one needs to obtain groups of mutually QWC terms. In the graph 
representation, this means finding fully-connected sub-graphs or {\it cliques}. To optimize the 
measurement process we are interested in the minimum number of cliques (Fig.~\ref{fig:QWCG}, middle panel)
\bea\notag
\hat H &=& \hat A_1 + \hat A_2 \\
\hat A_1 &=&\hat z_1 + \hat z_1\hat z_2 + \hat z_1\hat z_2\hat z_3 + \hat z_1\hat z_2\hat z_3\hat z_4 \\ 
\hat A_2 &=& \hat x_3\hat x_4 + \hat y_1\hat x_3\hat x_4 + \hat y_1\hat y_2\hat x_3\hat x_4.
\eea
It is easy to see that there are other solutions to the clique cover problem 
(Fig.~\ref{fig:QWCG}, lower panel)
\bea
\hat H &=& \hat A_1' +\hat A_2' + \hat A_3' \\
\hat A_1' &=& \hat z_1\hat z_2 + \hat z_1\hat z_2 \hat z_3 + \hat z_1\hat z_2 \hat z_3\hat z_4 \\
\hat A_2' &=& \hat z_1 + \hat x_3\hat x_4 \\ 
\hat A_3' &=& \hat y_1\hat x_3\hat x_4 + \hat y_1\hat y_2\hat x_3\hat x_4.
\eea    
This solution contains larger number of cliques and thus is non-optimal. 

The problem of finding the minimum number of cliques covering the graph is known 
as the minimum clique cover (MCC) problem. It is NP-hard in general, and its decision 
version is NP-complete. \cite{Karp:1972}
Therefore, we will focus on approximate polynomial approaches to MCC. 

\subsection{Solving the minimum clique cover problem}
\label{sec:MCC}

Two approaches to solving the MCC problem are considered in this work: 
1) through mapping to a graph coloring problem and
2) through employing a maximum clique search and removal. 
Both approaches involve solving NP-hard problems but they give rise to several intuitive heuristics. 
Here we present short rational for the connection of MCC to the two approaches
which are followed by descriptions of heuristics.

\subsubsection{Graph coloring} 

\begin{figure}[h!]
  \centering %
  \includegraphics[width=0.7\columnwidth]{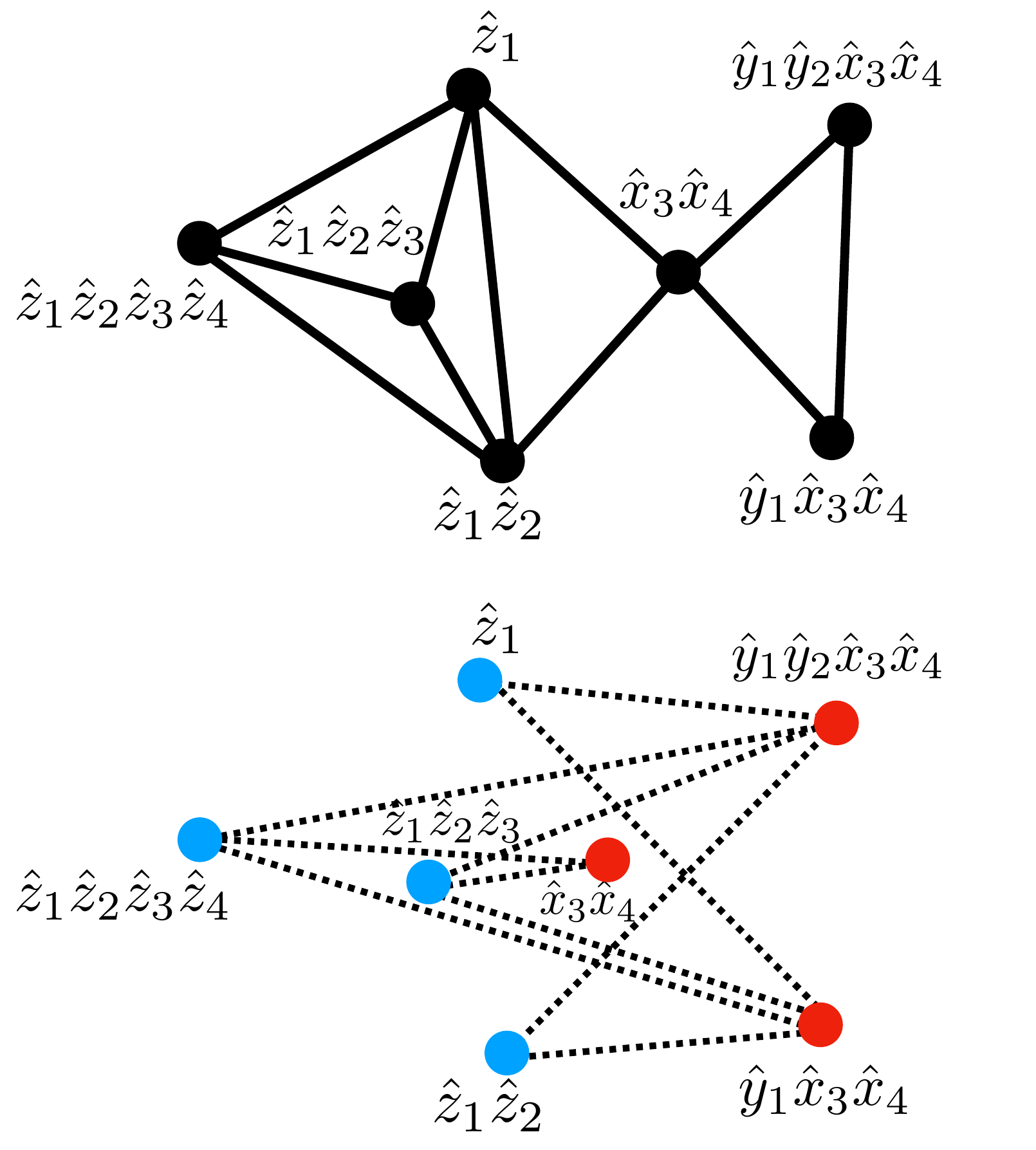}
  \caption{Graph representation of QWC terms in the Hamiltonian \eq{eq:Hex} (upper panel), 
  the complementary graph and its coloring (lower panel), the chromatic number is 2 which corresponds 
  to 2 cliques.}
  \label{fig:color}
\end{figure}

The MCC problem for a given graph $G$ can be mapped to a coloring problem 
for a complement graph $\bar G$ (see \fig{fig:color}). 
The coloring problem searches for the minimum number of colors that are needed to color $\bar G$  vertices 
so that any two connected vertices do not share the same color. The minimum number of colors 
(i.e. the chromatic number $\chi(\bar G)$) is 
equal to the minimum number of cliques in $G$. Each clique corresponds to vertices of $\bar G$ colored 
with the same color. Indeed, all $\bar G$ vertices of the same color are disconnected and hence form a clique in $G$. Proving that the color based cover is optimal can be easily done by leading to contradiction. 
Assuming that there is a clique cover that has a fewer number of cliques than $\chi(\bar G)$, 
one can obtain coloring of $\bar{G}$ that contains fewer colors than $\chi(\bar G)$ by coloring all vertices 
of a single clique in one color because they are disconnected in $\bar{G}$. 

To solve the graph coloring problem, the sequential vertex coloring algorithm is applied, which proceeds as follows. 
Given the ordering of vertices $v_1, v_2, ..., v_n$, vertex $v_1$ is colored with color 1, $k = 1$. 
To color a subsequent vertex, a set of colors of its neighbors is considered. 
If the set contains all $k$ colors, then color $k+1$ 
is assigned to the vertex, and $k$ is increased by 1. In case if not all $k$ colors are present in the 
set of colors for neighbors, the lowest available color is chosen to color the vertex. 

This algorithm does not produce the minimum number of colors, 
however, its complexity is polynominal with the number of graph vertices. 
The resulting number of colors $k$ depends on the ordering of vertices in the algorithm input, 
which led to development of heuristics to produce a lower number of colors.\cite{DeWerra:1990} 
We tested seven ordering heuristics found in previous works:

\paragraph*{Greedy Coloring (GC):} This algorithm uses the sequence of vertices corresponding to
OpenFermion ordering of Pauli words in qubit Hamiltonian generation. 
We label this algorithm GC because it is equivalent in formulation to the ``greedy'' procedure in the 
{$\texttt{group\_experiment}$} pyQuil routine.\cite{Rigetti_doc}

\paragraph*{Largest First (LF):} This algorithm puts the vertices of $\bar{G}$
in the non-increasing degree order.\cite{Welsh:1967}

\paragraph*{Smallest Last (SL):} The vertex with the smallest degree in $\bar{G}$ is 
 placed at the end of the list as $v_n$, where $n$ is the number of vertices in $\bar{G}$. 
A vertex at the $i^{\rm th}$ position in the list is the one with the smallest degree in the graph 
$\bar{G} - \{v_n, v_{(n-1)}, ..., v_{(i+1)}\}$.\cite{Matula:1972}

\paragraph*{DSATUR:} The largest degree vertex of $\bar{G}$ is assigned a color first. 
Then the order is established dynamically by coloring the vertex that is adjacent to the largest number of 
colored vertices (such a vertex is referred to as the most {\it saturated}).\cite{Brelaz:1973}

\paragraph*{Recursive Largest First (RLF):} The vertex with the largest degree in $\bar{G}$ is colored with color 1. 
The set of uncolored vertices is split into two subsets: $N_0$ including vertices that are not adjacent to any colored 
vertex and $N_1$ containing the rest. Then, in $N_0$ the vertex with the largest number of neighbors from 
$N_1$ is colored with the current color. $N_0$ and $N_1$ are updated.
This continues until all elements of $N_0$ are colored. When $N_0$ is exhausted, current color is increased and 
the process repeats {\BC for the rest of the graph that is uncolored}.\cite{Leighton:1979}

\paragraph*{Dutton and Brigham technique (DB):} Among disconnected pairs of vertices in $\bar{G}$, the pair with
 the biggest number of common neighbors is chosen. Then two vertices are merged in one which is connected 
 to the neighbors of both. The process is repeated until a clique is produced, with each vertex representing a 
 separate color. The vertices that were merged into one vertex of the produced clique are assigned the color of 
 this vertex.\cite{dutton_brigham_1981}

\paragraph*{COSINE:} This is variation of the DB scheme, where the first pair of disconnected vertices 
is chosen and merged as in DB. The next pair is made of the vertex 
obtained during the merger and a disconnected vertex which has the biggest number of common neighbors
with the first vertex in the pair. When there is no vertex that is disconnected from the current one, a new pair is chosen. Merging of pairs repeats until there are no more disconnected vertices. 
This process produces a clique and vertices that were merged into one vertex of the produced clique 
are assigned the color of the merged vertex.\cite{hertz_1990} 

{\BC Formally, GC and LF have $O(e+n)$ computational scaling, where $n$ and $e$ are the number of vertices and edges
 in $\bar{G}$, respectively. 
Vertices of the graphs originating from the QWC property of molecular Hamiltonians have relatively low degrees, therefore, those of 
their complementary graphs are quite high and practically $e\sim n^2$.} 
{\BC In general, RLF has cubic complexity in $n$, while for graphs with $k\cdot e \sim n^2$, its complexity is 
quadratic in $n$.\cite{Leighton:1979} 
For graphs studied in this work, $k\cdot e \sim n^3$, and thus, RLF's complexity is $O(n^3)$. }
DSATUR, DB, and COSINE have $O(n^3)$ computational scaling.

\subsubsection{Maximum clique search and removal}  

\begin{figure}[h!]
  \centering %
  \includegraphics[width=1\columnwidth]{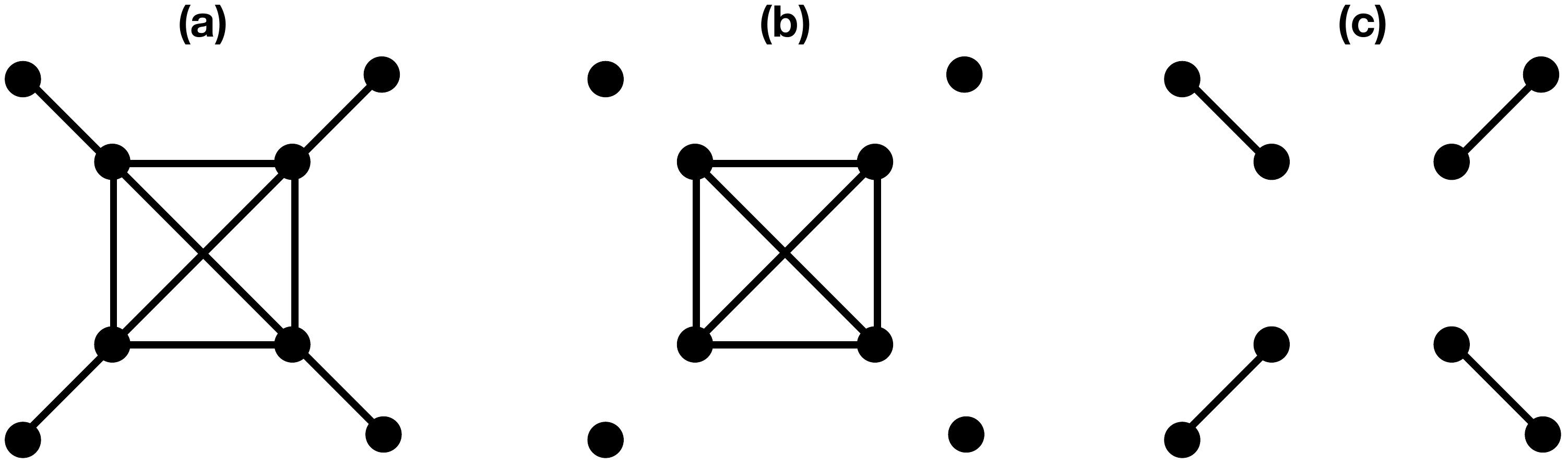}
  \caption{A graph example (a) where the maximum clique search and removal 
  produces a larger clique cover (b) than the minimum clique cover (c).}
  \label{fig:MCS}
\end{figure}

Repeating steps of searching for the maximum clique and removing it from the graph 
gives an approximate solution for MCC. Figure~\ref{fig:MCS} illustrates that even if the maximum clique
search is done exactly, which is an NP-hard problem, the obtained solution does not necessarily form 
the minimum clique cover. 
The first heuristic in this category uses an improvement on the Bron-Kerbosch the maximum clique search 
algorithm by Tomita {\it et al.}\cite{tomita_tanaka_takahashi_2006}, which is available as 
the {$\texttt{FindClique}$} procedure in Wolfram Mathematica.\cite{mathematica} 
Another heuristic algorithm is the polynomial Ramsey algorithm,\cite{boppana} 
implemented in the Python NetworkX library and used in the {$\texttt{group\_experiment}$} pyQuil 
routine\cite{Rigetti_doc}. We will refer to these two algorithms as BKT and Ramsey, respectively.


\begin{table*}[!htbp]
   \caption{The total number of Hamiltonian terms (Total) and the number of QWC groups produced by different heuristics for systems 
   with up to 1100 terms. The STO-3G basis has been used for all Hamiltonians unless specified otherwise.
   BK and JW denote the Bravyi-Kitaev and Jordan-Wigner fermion-qubit transformations.}
  \label{tab:var}
  \centering
    \begin{ruledtabular}
   \begin{tabular}{@{}lccccccccccc@{}}
    Systems & $N$ & Total & GC & LF & SL & DSATUR & RLF & DB & COSINE & Ramsey & BKT\\
    \hline
    H$_2$ (BK)& 4 & 15 & 3 & 3  & 3 & 3 &  3  & 3 & 3  & 3 & 3\\
    LiH  (Parity)& 4 & 100 & 25 & 25 & 25 & 25 & 25 & 25 & 25 & 25 & 25\\    
    H$_2$O (6-31G, BK)& 6 & 165 & 36 & 34 & 34 & 34 & 34 & 34 & 34 & 38 & 34 \\
    BeH$_2$  (BK) & 14 & 666 & 175 & 172 & 172 & 172 & 172 & 172 & 176 & 190 & 175\\
BeH$_2$  (JW)& 14 & 666 & 218 & 208 & 204 & 210 & 203 & 208 & 211& 225 & 216\\
H$_2$O  (BK) & 14 & 1086 & 320 & 313 & 316 & 315 & 311 & 313 & 308 & 322 & 319\\
H$_2$O (JW) & 14 & 1086 & 355 & 322 & 322 & 329 & 322 & 326 & 331 & 360 & 348\\
     \end{tabular}
       \end{ruledtabular}
\end{table*}

\section{Numerical studies and discussion}
\label{sec:numer-stud-disc}

To assess the heuristics we apply them to a set of small molecule Hamiltonians 
obtained using the STO-3G and 6-31G bases (see Tables~\ref{tab:var} and \ref{tab:varst}). 
Details of generating these Hamiltonians are given in Supplementary Material. Some 
of these systems were used previously to illustrate performance of quantum computing techniques.
\cite{Kandala:2017/nature/242,Hempel:2018/prx/031022,Ryabinkin:2018/qcc}.

\begin{table*}[!htbp]
   \caption{Comparison of {\BC LF} results for Bravyi-Kitaev (BK) and Jordan-Wigner (JW) transformed Hamiltonians: the number of cliques (Cliques), their maximum size (Max Size) and standard deviation of their size distribution (STD). The total number of Hamiltonian terms (Total) is almost everywhere the same for JW and BK; for the last two systems, JW numbers are in parenthesis.}
  \label{tab:varst}
    \begin{ruledtabular}
   \begin{tabular}{@{}lcccccccc@{}}
 \multirow{2}{*}{Systems} & \multirow{2}{*}{$N$}& \multirow{2}{*}{Total} & \multicolumn{3}{c}{BK} & \multicolumn{3}{c}{JW}\\
 \cline{4-6}\cline{7-9}
	 & & & Cliques & Max Size & STD & Cliques &Max Size& STD \\
    \hline
     BeH$_2$ / STO-3G & 14 & 666 &  172 &  22 & 3.7 & 208 & 57 & 5.1\\
	H$_2$O / STO-3G & 14 & 1086 & 313 & 21 & 3.3 & 322 &  57 & 4.4\\
	NH$_3$ / STO-3G & 16 & 3609 & 1271 & 26 & 2.1 & 1201 & 80 & 3.3\\
	N$_2$ / STO-3G & 20 & 2951 &1178 & 36 & 2.9 & 1187 & 138 & 4.8 \\
    BeH$_2$ / 6-31G & 26 & 9204 & 2983 & 110 & 3.9 & 2720 &255 & 5.8\\
H$_2$O / 6-31G & 26 & 12732 & 3878 &  81 & 3.8 & 3719 & 255 & 5.4 \\
    NH$_3$ / 6-31G & 30 & 52758  (52806) & 14924 & 73 & 3.3 & 14907 &353 & 4.1 \\
    N$_2$ / 6-31G & 36 & 34639 (34583) & 11538 &  114 & 3.6 & 12399 & 530 & 5.5 \\
     \end{tabular}
       \end{ruledtabular}
\end{table*}

Table~\ref{tab:var} summarizes results of the QWC partitioning obtained using different heuristics. 
The number of $\hat A_n$ groups is 3 to 5 times fewer than the number of Pauli words in qubit Hamiltonians.
To test heuristics further, sizes of the first three Hamiltonians allowed us to examine up to 100,000 different randomly 
generated vertex orderings for graph coloring algorithms. There were no fewer clique cover numbers found for those 
orderings than the minimal ones reported in the Table~\ref{tab:var}. 



On average, the difference between various methods does not exceed approximately 10\% of the total 
number of cliques (see Table~\ref{tab:var}). 
Therefore, it is reasonable to use an approach that has the lowest computational cost and performs well overall. 
{\BC LF} performance was superior to the other techniques in both the number 
of produced cliques and execution time. Thus we used {\BC LF} for solution of the MCC problem for larger Hamiltonians and 
for exploring clique statistics (see Table~\ref{tab:varst}). Even though all heuristics except BKT are polynomial in scaling, 
DB, COSINE, and Ramsey spent almost two orders of magnitude longer times for 14-qubit H$_2$O than other algorithms, and therefore could not be recommended for use in larger Hamiltonians. {\BC The only other scheme that sometimes outperforms LF and  
may not be too computationally expensive is RLF.} In all systems larger than 6 qubits, the GC algorithm 
that uses an OpenFermion ordering of Pauli words can be easily improved by switching to other orders (e.g. LF, SL, or LRF).      

Application of the {\BC LF} heuristic to larger Hamiltonians in Table~\ref{tab:varst} demonstrates 
higher than or equal to three-fold reduction of the number of cliques compare to the 
total number of terms independent of the type of the fermion-qubit mapping. The main difference between cliques in the 
JW and BK transformations is in distributions of their sizes. Maximum size $\hat A_n$ groups for the JW transformed Hamiltonians 
is almost three times greater in size than those for the BK transformed Hamiltonians. 
Similar trend can be observed for the standard deviations of clique sizes with an approximate ratio of one and a half 
between JW and BK transformed Hamiltonians.

\section{Conclusions}
\label{sec:conclusions}

We have introduced and studied a new method for partitioning of the qubit Hamiltonian
in the VQE approach to the electronic structure problem. The main idea of our approach is 
to represent the Hamiltonian as a graph where every vertex corresponds to a single Pauli word and 
the edges are connecting the terms that are qubit-wise commuting. In this representation,
the problem of grouping terms that can be measured simultaneously by single-qubit measurement
is equivalent to finding a fully connected subgraphs (cliques). To obtain optimal partitioning 
the number of groups should be the fewest. This is well-known problem in discrete math, 
the minimum clique cover problem. We benchmarked few heuristic polynomial algorithms 
to approximately solve this NP-hard problem and found that the number of qubit-wise commuting 
groups can be reduced three times from the total number of the Hamiltonian terms. 

The difference in numbers of groups produced by different algorithms did not exceed 10\%, 
therefore fastest algorithms, {\BC LF and RLF, are generally recommended.}
{\BC Considering the growth of the number of Hamiltonian terms, $O(N^4)$, with the number of qubits $N$, 
the scalings of LF and RLF algorithms are $O(N^8)$ and $O(N^{12})$, respectively.}
No significant difference in the number of groups was observed between corresponding 
Hamiltonians obtained by the BK and JW transformations.
{\BC Numerical simulations revealed that the} number of qubit-wise commuting groups 
is proportional to the number of Pauli words which grows as $O(N^4)$. 
Thus, to treat large systems one needs to introduce further improvements in the measurement preprocessing.


Since the submission of this paper to arXiv, several new proposals addressing the measurement problem were put forward.\cite{MoscaA,ThomsonA,Izmaylov2019UP,BabbushA,ChicagoA}
 The main difference between the current approach and the subsequent ones is the complexity of pre-measurement unitary transformations. 
 The current approach involves only one-qubit unitary rotations, while those in Refs.~\cite{MoscaA,ThomsonA,Izmaylov2019UP,BabbushA,ChicagoA} employ multi-qubit transformations. 
 This makes the current approach especially well suited for low-depth circuits. As shown in Refs.~\cite{ThomsonA,BabbushA,Izmaylov2019UP}, 
 multi-qubit transformations can reduce the scaling of the number of measurable groups, but for the price 
 of introducing additional multi-qubit gates.

\section*{Supplementary Material}
Details of generating the Hamiltonians for the systems presented in Tables~\ref{tab:var} and \ref{tab:varst}
are given in Supplementary Material. 

\section*{Acknowledgement}
A.F.I. is grateful to Alan Aspuru-Guzik, Max Radin, Nicholas Rubin, and Matthew Harrigan 
 for useful discussions and helpful suggestions.  
A.F.I. acknowledges financial support from Zapata Computing, Inc., the Natural Sciences and
Engineering Research Council of Canada, the Mitacs Globalink Program, and 
the Google Quantum Research Program. 


\begin{thebibliography}{34}%
\makeatletter
\providecommand \@ifxundefined [1]{%
 \@ifx{#1\undefined}
}%
\providecommand \@ifnum [1]{%
 \ifnum #1\expandafter \@firstoftwo
 \else \expandafter \@secondoftwo
 \fi
}%
\providecommand \@ifx [1]{%
 \ifx #1\expandafter \@firstoftwo
 \else \expandafter \@secondoftwo
 \fi
}%
\providecommand \natexlab [1]{#1}%
\providecommand \enquote  [1]{``#1''}%
\providecommand \bibnamefont  [1]{#1}%
\providecommand \bibfnamefont [1]{#1}%
\providecommand \citenamefont [1]{#1}%
\providecommand \href@noop [0]{\@secondoftwo}%
\providecommand \href [0]{\begingroup \@sanitize@url \@href}%
\providecommand \@href[1]{\@@startlink{#1}\@@href}%
\providecommand \@@href[1]{\endgroup#1\@@endlink}%
\providecommand \@sanitize@url [0]{\catcode `\\12\catcode `\$12\catcode
  `\&12\catcode `\#12\catcode `\^12\catcode `\_12\catcode `\%12\relax}%
\providecommand \@@startlink[1]{}%
\providecommand \@@endlink[0]{}%
\providecommand \url  [0]{\begingroup\@sanitize@url \@url }%
\providecommand \@url [1]{\endgroup\@href {#1}{\urlprefix }}%
\providecommand \urlprefix  [0]{URL }%
\providecommand \Eprint [0]{\href }%
\providecommand \doibase [0]{http://dx.doi.org/}%
\providecommand \selectlanguage [0]{\@gobble}%
\providecommand \bibinfo  [0]{\@secondoftwo}%
\providecommand \bibfield  [0]{\@secondoftwo}%
\providecommand \translation [1]{[#1]}%
\providecommand \BibitemOpen [0]{}%
\providecommand \bibitemStop [0]{}%
\providecommand \bibitemNoStop [0]{.\EOS\space}%
\providecommand \EOS [0]{\spacefactor3000\relax}%
\providecommand \BibitemShut  [1]{\csname bibitem#1\endcsname}%
\let\auto@bib@innerbib\@empty
\bibitem [{\citenamefont {Yung}\ \emph {et~al.}(2014)\citenamefont {Yung},
  \citenamefont {Casanova}, \citenamefont {Mezzacapo}, \citenamefont {McClean},
  \citenamefont {Lamata}, \citenamefont {Aspuru-Guzik},\ and\ \citenamefont
  {Solano}}]{Yung:2014iv}%
  \BibitemOpen
  \bibfield  {author} {\bibinfo {author} {\bibfnamefont {M.~H.}\ \bibnamefont
  {Yung}}, \bibinfo {author} {\bibfnamefont {J.}~\bibnamefont {Casanova}},
  \bibinfo {author} {\bibfnamefont {A.}~\bibnamefont {Mezzacapo}}, \bibinfo
  {author} {\bibfnamefont {J.}~\bibnamefont {McClean}}, \bibinfo {author}
  {\bibfnamefont {L.}~\bibnamefont {Lamata}}, \bibinfo {author} {\bibfnamefont
  {A.}~\bibnamefont {Aspuru-Guzik}}, \ and\ \bibinfo {author} {\bibfnamefont
  {E.}~\bibnamefont {Solano}},\ }\href@noop {} {\bibfield  {journal} {\bibinfo
  {journal} {Sci. Rep.}\ }\textbf {\bibinfo {volume} {4}},\ \bibinfo {pages}
  {714} (\bibinfo {year} {2014})}\BibitemShut {NoStop}%
\bibitem [{\citenamefont {Peruzzo}\ \emph {et~al.}(2014)\citenamefont
  {Peruzzo}, \citenamefont {McClean}, \citenamefont {Shadbolt}, \citenamefont
  {Yung}, \citenamefont {Zhou}, \citenamefont {Love}, \citenamefont
  {Aspuru-Guzik},\ and\ \citenamefont {O'Brien}}]{Peruzzo:2014/ncomm/4213}%
  \BibitemOpen
  \bibfield  {author} {\bibinfo {author} {\bibfnamefont {A.}~\bibnamefont
  {Peruzzo}}, \bibinfo {author} {\bibfnamefont {J.}~\bibnamefont {McClean}},
  \bibinfo {author} {\bibfnamefont {P.}~\bibnamefont {Shadbolt}}, \bibinfo
  {author} {\bibfnamefont {M.-H.}\ \bibnamefont {Yung}}, \bibinfo {author}
  {\bibfnamefont {X.-Q.}\ \bibnamefont {Zhou}}, \bibinfo {author}
  {\bibfnamefont {P.~J.}\ \bibnamefont {Love}}, \bibinfo {author}
  {\bibfnamefont {A.}~\bibnamefont {Aspuru-Guzik}}, \ and\ \bibinfo {author}
  {\bibfnamefont {J.~L.}\ \bibnamefont {O'Brien}},\ }\href@noop {} {\bibfield
  {journal} {\bibinfo  {journal} {Nat. Commun.}\ }\textbf {\bibinfo {volume}
  {5}},\ \bibinfo {pages} {4213} (\bibinfo {year} {2014})}\BibitemShut
  {NoStop}%
\bibitem [{\citenamefont {McClean}\ \emph {et~al.}(2016)\citenamefont
  {McClean}, \citenamefont {Romero}, \citenamefont {Babbush},\ and\
  \citenamefont {Aspuru-Guzik}}]{Jarrod:2016/njp/023023}%
  \BibitemOpen
  \bibfield  {author} {\bibinfo {author} {\bibfnamefont {J.~R.}\ \bibnamefont
  {McClean}}, \bibinfo {author} {\bibfnamefont {J.}~\bibnamefont {Romero}},
  \bibinfo {author} {\bibfnamefont {R.}~\bibnamefont {Babbush}}, \ and\
  \bibinfo {author} {\bibfnamefont {A.}~\bibnamefont {Aspuru-Guzik}},\ }\href
  {\doibase 10.1088/1367-2630/18/2/023023} {\bibfield  {journal} {\bibinfo
  {journal} {N. J. Phys.}\ }\textbf {\bibinfo {volume} {18}},\ \bibinfo {pages}
  {023023} (\bibinfo {year} {2016})}\BibitemShut {NoStop}%
\bibitem [{\citenamefont {Wecker}\ \emph {et~al.}(2015)\citenamefont {Wecker},
  \citenamefont {Hastings},\ and\ \citenamefont
  {Troyer}}]{Wecker:2015/pra/042303}%
  \BibitemOpen
  \bibfield  {author} {\bibinfo {author} {\bibfnamefont {D.}~\bibnamefont
  {Wecker}}, \bibinfo {author} {\bibfnamefont {M.~B.}\ \bibnamefont
  {Hastings}}, \ and\ \bibinfo {author} {\bibfnamefont {M.}~\bibnamefont
  {Troyer}},\ }\href {\doibase 10.1103/PhysRevA.92.042303} {\bibfield
  {journal} {\bibinfo  {journal} {Phys. Rev. A}\ }\textbf {\bibinfo {volume}
  {92}},\ \bibinfo {pages} {042303} (\bibinfo {year} {2015})}\BibitemShut
  {NoStop}%
\bibitem [{\citenamefont {Olson}\ \emph {et~al.}(2017)\citenamefont {Olson},
  \citenamefont {Cao}, \citenamefont {Romero}, \citenamefont {Johnson},
  \citenamefont {Dallaire-Demers}, \citenamefont {Sawaya}, \citenamefont
  {Narang}, \citenamefont {Kivlichan}, \citenamefont {Wasielewski},\ and\
  \citenamefont {Aspuru-Guzik}}]{Olson:2017ud}%
  \BibitemOpen
  \bibfield  {author} {\bibinfo {author} {\bibfnamefont {J.}~\bibnamefont
  {Olson}}, \bibinfo {author} {\bibfnamefont {Y.}~\bibnamefont {Cao}}, \bibinfo
  {author} {\bibfnamefont {J.}~\bibnamefont {Romero}}, \bibinfo {author}
  {\bibfnamefont {P.}~\bibnamefont {Johnson}}, \bibinfo {author} {\bibfnamefont
  {P.-L.}\ \bibnamefont {Dallaire-Demers}}, \bibinfo {author} {\bibfnamefont
  {N.}~\bibnamefont {Sawaya}}, \bibinfo {author} {\bibfnamefont
  {P.}~\bibnamefont {Narang}}, \bibinfo {author} {\bibfnamefont
  {I.}~\bibnamefont {Kivlichan}}, \bibinfo {author} {\bibfnamefont
  {M.}~\bibnamefont {Wasielewski}}, \ and\ \bibinfo {author} {\bibfnamefont
  {A.}~\bibnamefont {Aspuru-Guzik}},\ }\href@noop {} {\bibfield  {journal}
  {\bibinfo  {journal} {arXiv.org}\ } (\bibinfo {year} {2017})},\ \Eprint
  {http://arxiv.org/abs/1706.05413v2} {1706.05413v2} \BibitemShut {NoStop}%
\bibitem [{\citenamefont {McArdle}\ \emph {et~al.}(2018)\citenamefont
  {McArdle}, \citenamefont {Endo}, \citenamefont {Aspuru-Guzik}, \citenamefont
  {Benjamin},\ and\ \citenamefont {Yuan}}]{McArdle:2018we}%
  \BibitemOpen
  \bibfield  {author} {\bibinfo {author} {\bibfnamefont {S.}~\bibnamefont
  {McArdle}}, \bibinfo {author} {\bibfnamefont {S.}~\bibnamefont {Endo}},
  \bibinfo {author} {\bibfnamefont {A.}~\bibnamefont {Aspuru-Guzik}}, \bibinfo
  {author} {\bibfnamefont {S.}~\bibnamefont {Benjamin}}, \ and\ \bibinfo
  {author} {\bibfnamefont {X.}~\bibnamefont {Yuan}},\ }\href@noop {} {\bibfield
   {journal} {\bibinfo  {journal} {arXiv.org}\ } (\bibinfo {year} {2018})},\
  \Eprint {http://arxiv.org/abs/1808.10402v1} {1808.10402v1} \BibitemShut
  {NoStop}%
\bibitem [{\citenamefont {Jordan}\ and\ \citenamefont
  {Wigner}(1928)}]{Jordan:1928/zphys/631}%
  \BibitemOpen
  \bibfield  {author} {\bibinfo {author} {\bibfnamefont {P.}~\bibnamefont
  {Jordan}}\ and\ \bibinfo {author} {\bibfnamefont {E.}~\bibnamefont
  {Wigner}},\ }\href {\doibase 10.1007/BF01331938} {\bibfield  {journal}
  {\bibinfo  {journal} {Z. Phys.}\ }\textbf {\bibinfo {volume} {47}},\ \bibinfo
  {pages} {631} (\bibinfo {year} {1928})}\BibitemShut {NoStop}%
\bibitem [{\citenamefont {Bravyi}\ and\ \citenamefont
  {Kitaev}(2002)}]{Bravyi:2002/aph/210}%
  \BibitemOpen
  \bibfield  {author} {\bibinfo {author} {\bibfnamefont {S.~B.}\ \bibnamefont
  {Bravyi}}\ and\ \bibinfo {author} {\bibfnamefont {A.~Y.}\ \bibnamefont
  {Kitaev}},\ }\href {\doibase 10.1006/aphy.2002.6254} {\bibfield  {journal}
  {\bibinfo  {journal} {Ann. Phys.}\ }\textbf {\bibinfo {volume} {298}},\
  \bibinfo {pages} {210} (\bibinfo {year} {2002})}\BibitemShut {NoStop}%
\bibitem [{\citenamefont {Seeley}\ \emph {et~al.}(2012)\citenamefont {Seeley},
  \citenamefont {Richard},\ and\ \citenamefont
  {Love}}]{Seeley:2012/jcp/224109}%
  \BibitemOpen
  \bibfield  {author} {\bibinfo {author} {\bibfnamefont {J.~T.}\ \bibnamefont
  {Seeley}}, \bibinfo {author} {\bibfnamefont {M.~J.}\ \bibnamefont {Richard}},
  \ and\ \bibinfo {author} {\bibfnamefont {P.~J.}\ \bibnamefont {Love}},\
  }\href {\doibase 10.1063/1.4768229} {\bibfield  {journal} {\bibinfo
  {journal} {J. Chem. Phys.}\ }\textbf {\bibinfo {volume} {137}},\ \bibinfo
  {pages} {224109} (\bibinfo {year} {2012})}\BibitemShut {NoStop}%
\bibitem [{\citenamefont {Tranter}\ \emph {et~al.}(2015)\citenamefont
  {Tranter}, \citenamefont {Sofia}, \citenamefont {Seeley}, \citenamefont
  {Kaicher}, \citenamefont {McClean}, \citenamefont {Babbush}, \citenamefont
  {Coveney}, \citenamefont {Mintert}, \citenamefont {Wilhelm},\ and\
  \citenamefont {Love}}]{Tranter:2015/ijqc/1431}%
  \BibitemOpen
  \bibfield  {author} {\bibinfo {author} {\bibfnamefont {A.}~\bibnamefont
  {Tranter}}, \bibinfo {author} {\bibfnamefont {S.}~\bibnamefont {Sofia}},
  \bibinfo {author} {\bibfnamefont {J.}~\bibnamefont {Seeley}}, \bibinfo
  {author} {\bibfnamefont {M.}~\bibnamefont {Kaicher}}, \bibinfo {author}
  {\bibfnamefont {J.}~\bibnamefont {McClean}}, \bibinfo {author} {\bibfnamefont
  {R.}~\bibnamefont {Babbush}}, \bibinfo {author} {\bibfnamefont {P.~V.}\
  \bibnamefont {Coveney}}, \bibinfo {author} {\bibfnamefont {F.}~\bibnamefont
  {Mintert}}, \bibinfo {author} {\bibfnamefont {F.}~\bibnamefont {Wilhelm}}, \
  and\ \bibinfo {author} {\bibfnamefont {P.~J.}\ \bibnamefont {Love}},\ }\href
  {\doibase 10.1002/qua.24969} {\bibfield  {journal} {\bibinfo  {journal} {Int.
  J. Quantum Chem.}\ }\textbf {\bibinfo {volume} {115}},\ \bibinfo {pages}
  {1431} (\bibinfo {year} {2015})}\BibitemShut {NoStop}%
\bibitem [{\citenamefont {Setia}\ and\ \citenamefont
  {Whitfield}(2018)}]{Setia:2017/ArXiv/1712.00446}%
  \BibitemOpen
  \bibfield  {author} {\bibinfo {author} {\bibfnamefont {K.}~\bibnamefont
  {Setia}}\ and\ \bibinfo {author} {\bibfnamefont {J.~D.}\ \bibnamefont
  {Whitfield}},\ }\href {\doibase 10.1063/1.5019371} {\bibfield  {journal}
  {\bibinfo  {journal} {J. Chem. Phys.}\ }\textbf {\bibinfo {volume} {148}},\
  \bibinfo {pages} {164104} (\bibinfo {year} {2018})}\BibitemShut {NoStop}%
\bibitem [{\citenamefont {Havl{\'{i}}{\v{c}}ek}\ \emph
  {et~al.}(2017)\citenamefont {Havl{\'{i}}{\v{c}}ek}, \citenamefont {Troyer},\
  and\ \citenamefont {Whitfield}}]{Havlicek:2017/pra/032332}%
  \BibitemOpen
  \bibfield  {author} {\bibinfo {author} {\bibfnamefont {V.}~\bibnamefont
  {Havl{\'{i}}{\v{c}}ek}}, \bibinfo {author} {\bibfnamefont {M.}~\bibnamefont
  {Troyer}}, \ and\ \bibinfo {author} {\bibfnamefont {J.~D.}\ \bibnamefont
  {Whitfield}},\ }\href {\doibase 10.1103/PhysRevA.95.032332} {\bibfield
  {journal} {\bibinfo  {journal} {Phys. Rev. A}\ }\textbf {\bibinfo {volume}
  {95}},\ \bibinfo {pages} {032332} (\bibinfo {year} {2017})}\BibitemShut
  {NoStop}%
\bibitem [{\citenamefont {Cirac}\ and\ \citenamefont
  {Zoller}(2012)}]{cirac:2012}%
  \BibitemOpen
  \bibfield  {author} {\bibinfo {author} {\bibfnamefont {J.~I.}\ \bibnamefont
  {Cirac}}\ and\ \bibinfo {author} {\bibfnamefont {P.}~\bibnamefont {Zoller}},\
  }\href {\doibase 10.1038/nphys2275} {\bibfield  {journal} {\bibinfo
  {journal} {Nat. Phys.}\ }\textbf {\bibinfo {volume} {8}},\ \bibinfo {pages}
  {264} (\bibinfo {year} {2012})}\BibitemShut {NoStop}%
\bibitem [{\citenamefont {Argüello-Luengo}\ \emph {et~al.}(2018)\citenamefont
  {Argüello-Luengo}, \citenamefont {González-Tudela}, \citenamefont {Shi},
  \citenamefont {Zoller},\ and\ \citenamefont {Cirac}}]{cirac:2018}%
  \BibitemOpen
  \bibfield  {author} {\bibinfo {author} {\bibfnamefont {J.}~\bibnamefont
  {Argüello-Luengo}}, \bibinfo {author} {\bibfnamefont {A.}~\bibnamefont
  {González-Tudela}}, \bibinfo {author} {\bibfnamefont {T.}~\bibnamefont
  {Shi}}, \bibinfo {author} {\bibfnamefont {P.}~\bibnamefont {Zoller}}, \ and\
  \bibinfo {author} {\bibfnamefont {J.~I.}\ \bibnamefont {Cirac}},\ }\href@noop
  {} {\bibfield  {journal} {\bibinfo  {journal} {arXiv.org}\ } (\bibinfo {year}
  {2018})},\ \Eprint {http://arxiv.org/abs/1807.09228} {1807.09228}
  \BibitemShut {NoStop}%
\bibitem [{\citenamefont {Kandala}\ \emph {et~al.}(2017)\citenamefont
  {Kandala}, \citenamefont {Mezzacapo}, \citenamefont {Temme}, \citenamefont
  {Takita}, \citenamefont {Brink}, \citenamefont {Chow},\ and\ \citenamefont
  {Gambetta}}]{Kandala:2017/nature/242}%
  \BibitemOpen
  \bibfield  {author} {\bibinfo {author} {\bibfnamefont {A.}~\bibnamefont
  {Kandala}}, \bibinfo {author} {\bibfnamefont {A.}~\bibnamefont {Mezzacapo}},
  \bibinfo {author} {\bibfnamefont {K.}~\bibnamefont {Temme}}, \bibinfo
  {author} {\bibfnamefont {M.}~\bibnamefont {Takita}}, \bibinfo {author}
  {\bibfnamefont {M.}~\bibnamefont {Brink}}, \bibinfo {author} {\bibfnamefont
  {J.~M.}\ \bibnamefont {Chow}}, \ and\ \bibinfo {author} {\bibfnamefont
  {J.~M.}\ \bibnamefont {Gambetta}},\ }\href {\doibase 10.1038/nature23879}
  {\bibfield  {journal} {\bibinfo  {journal} {Nature}\ }\textbf {\bibinfo
  {volume} {549}},\ \bibinfo {pages} {242} (\bibinfo {year}
  {2017})}\BibitemShut {NoStop}%
\bibitem [{\citenamefont {{Rigetti\ Computing}}(2018)}]{Rigetti_doc}%
  \BibitemOpen
  \bibfield  {author} {\bibinfo {author} {\bibnamefont {{Rigetti\
  Computing}}},\ }\href@noop {} {\enquote {\bibinfo {title} {{pyQuil 1.9}},}\ }
  (\bibinfo {year} {2018}),\ \bibinfo {note}
  {http://docs.rigetti.com/en/1.9/qpu.html}\BibitemShut {NoStop}%
\bibitem [{\citenamefont {Karp}(1972)}]{Karp:1972}%
  \BibitemOpen
  \bibfield  {author} {\bibinfo {author} {\bibfnamefont {R.~M.}\ \bibnamefont
  {Karp}},\ }in\ \href {\doibase 10.1007/978-1-4684-2001-2_9} {\emph {\bibinfo
  {booktitle} {Complexity of Computer Computations. The IBM Research Symposia
  Series}}},\ \bibinfo {editor} {edited by\ \bibinfo {editor} {\bibfnamefont
  {R.}~\bibnamefont {Miller}}, \bibinfo {editor} {\bibfnamefont
  {J.}~\bibnamefont {Thatcher}}, \ and\ \bibinfo {editor} {\bibfnamefont
  {J.}~\bibnamefont {Bohlinger}}}\ (\bibinfo  {publisher} {Springer, Boston,
  MA},\ \bibinfo {year} {1972})\ p.\ \bibinfo {pages} {85–103}\BibitemShut
  {NoStop}%
\bibitem [{\citenamefont {Werra}(1990)}]{DeWerra:1990}%
  \BibitemOpen
  \bibfield  {author} {\bibinfo {author} {\bibfnamefont {D.~D.}\ \bibnamefont
  {Werra}},\ }in\ \href {\doibase 10.1007/978-3-7091-9076-0_10} {\emph
  {\bibinfo {booktitle} {Computational Graph Theory. Computing
  Supplementum}}},\ Vol.~\bibinfo {volume} {7},\ \bibinfo {editor} {edited by\
  \bibinfo {editor} {\bibfnamefont {G.}~\bibnamefont {Tinhofer}}, \bibinfo
  {editor} {\bibfnamefont {E.}~\bibnamefont {Mayr}}, \bibinfo {editor}
  {\bibfnamefont {H.}~\bibnamefont {Noltemeier}}, \ and\ \bibinfo {editor}
  {\bibfnamefont {M.~M.}\ \bibnamefont {Syslo}}}\ (\bibinfo  {publisher}
  {Springer, Vienna},\ \bibinfo {year} {1990})\ pp.\ \bibinfo {pages}
  {191--208}\BibitemShut {NoStop}%
\bibitem [{\citenamefont {Welsh}(1967)}]{Welsh:1967}%
  \BibitemOpen
  \bibfield  {author} {\bibinfo {author} {\bibfnamefont {D.~J.~A.}\
  \bibnamefont {Welsh}},\ }\href {\doibase 10.1093/comjnl/10.1.85} {\bibfield
  {journal} {\bibinfo  {journal} {Comput. J.}\ }\textbf {\bibinfo {volume}
  {10}},\ \bibinfo {pages} {85–86} (\bibinfo {year} {1967})}\BibitemShut
  {NoStop}%
\bibitem [{\citenamefont {Matula}\ \emph {et~al.}(1972)\citenamefont {Matula},
  \citenamefont {Marble},\ and\ \citenamefont {Isaacson}}]{Matula:1972}%
  \BibitemOpen
  \bibfield  {author} {\bibinfo {author} {\bibfnamefont {D.~W.}\ \bibnamefont
  {Matula}}, \bibinfo {author} {\bibfnamefont {G.}~\bibnamefont {Marble}}, \
  and\ \bibinfo {author} {\bibfnamefont {J.~D.}\ \bibnamefont {Isaacson}},\
  }in\ \href {\doibase 10.1016/b978-1-4832-3187-7.50015-5} {\emph {\bibinfo
  {booktitle} {Graph Theory and Computing}}},\ \bibinfo {editor} {edited by\
  \bibinfo {editor} {\bibfnamefont {R.~C.}\ \bibnamefont {Read}}}\ (\bibinfo
  {publisher} {Academic Press},\ \bibinfo {year} {1972})\ pp.\ \bibinfo {pages}
  {109 -- 122}\BibitemShut {NoStop}%
\bibitem [{\citenamefont {Brélaz}(1979)}]{Brelaz:1973}%
  \BibitemOpen
  \bibfield  {author} {\bibinfo {author} {\bibfnamefont {D.}~\bibnamefont
  {Brélaz}},\ }\href {\doibase 10.1145/359094.359101} {\bibfield  {journal}
  {\bibinfo  {journal} {Commun. ACM}\ }\textbf {\bibinfo {volume} {22}},\
  \bibinfo {pages} {251–256} (\bibinfo {year} {1979})}\BibitemShut {NoStop}%
\bibitem [{\citenamefont {Leighton}(1979)}]{Leighton:1979}%
  \BibitemOpen
  \bibfield  {author} {\bibinfo {author} {\bibfnamefont {F.~T.}\ \bibnamefont
  {Leighton}},\ }\href@noop {} {\bibfield  {journal} {\bibinfo  {journal} {J.
  Res. Natl. Bur. Stand.}\ }\textbf {\bibinfo {volume} {84}},\ \bibinfo {pages}
  {489} (\bibinfo {year} {1979})}\BibitemShut {NoStop}%
\bibitem [{\citenamefont {Dutton}\ and\ \citenamefont
  {Brigham}(1981)}]{dutton_brigham_1981}%
  \BibitemOpen
  \bibfield  {author} {\bibinfo {author} {\bibfnamefont {R.~D.}\ \bibnamefont
  {Dutton}}\ and\ \bibinfo {author} {\bibfnamefont {R.~C.}\ \bibnamefont
  {Brigham}},\ }\href {\doibase 10.1093/comjnl/24.1.85} {\bibfield  {journal}
  {\bibinfo  {journal} {Comput. J.}\ }\textbf {\bibinfo {volume} {24}},\
  \bibinfo {pages} {85–86} (\bibinfo {year} {1981})}\BibitemShut {NoStop}%
\bibitem [{\citenamefont {Hertz}(1990)}]{hertz_1990}%
  \BibitemOpen
  \bibfield  {author} {\bibinfo {author} {\bibfnamefont {A.}~\bibnamefont
  {Hertz}},\ }\href {\doibase 10.1016/0095-8956(90)90078-e} {\bibfield
  {journal} {\bibinfo  {journal} {J. Comb. Theory}\ }\textbf {\bibinfo {volume}
  {50}},\ \bibinfo {pages} {231–240} (\bibinfo {year} {1990})}\BibitemShut
  {NoStop}%
\bibitem [{\citenamefont {Tomita}\ \emph {et~al.}(2006)\citenamefont {Tomita},
  \citenamefont {Tanaka},\ and\ \citenamefont
  {Takahashi}}]{tomita_tanaka_takahashi_2006}%
  \BibitemOpen
  \bibfield  {author} {\bibinfo {author} {\bibfnamefont {E.}~\bibnamefont
  {Tomita}}, \bibinfo {author} {\bibfnamefont {A.}~\bibnamefont {Tanaka}}, \
  and\ \bibinfo {author} {\bibfnamefont {H.}~\bibnamefont {Takahashi}},\ }\href
  {\doibase 10.1016/j.tcs.2006.06.015} {\bibfield  {journal} {\bibinfo
  {journal} {Theor. Comput. Sci.}\ }\textbf {\bibinfo {volume} {363}},\
  \bibinfo {pages} {28–42} (\bibinfo {year} {2006})}\BibitemShut {NoStop}%
\bibitem [{mat()}]{mathematica}%
  \BibitemOpen
  \href@noop {} {}\bibinfo {note} {Wolfram Research, Inc., Mathematica, Version
  12.0, Champaign, IL (2019)}\BibitemShut {NoStop}%
\bibitem [{\citenamefont {Boppana}\ and\ \citenamefont
  {Halldórsson}(1992)}]{boppana}%
  \BibitemOpen
  \bibfield  {author} {\bibinfo {author} {\bibfnamefont {R.}~\bibnamefont
  {Boppana}}\ and\ \bibinfo {author} {\bibfnamefont {M.~M.}\ \bibnamefont
  {Halldórsson}},\ }\href {\doibase 10.1007/bf01994876} {\bibfield  {journal}
  {\bibinfo  {journal} {BIT Numer. Math}\ }\textbf {\bibinfo {volume} {32}},\
  \bibinfo {pages} {180–196} (\bibinfo {year} {1992})}\BibitemShut {NoStop}%
\bibitem [{\citenamefont {Hempel}\ \emph {et~al.}(2018)\citenamefont {Hempel},
  \citenamefont {Maier}, \citenamefont {Romero}, \citenamefont {McClean},
  \citenamefont {Monz}, \citenamefont {Shen}, \citenamefont {Jurcevic},
  \citenamefont {Lanyon}, \citenamefont {Love}, \citenamefont {Babbush},
  \citenamefont {Aspuru-Guzik}, \citenamefont {Blatt},\ and\ \citenamefont
  {Roos}}]{Hempel:2018/prx/031022}%
  \BibitemOpen
  \bibfield  {author} {\bibinfo {author} {\bibfnamefont {C.}~\bibnamefont
  {Hempel}}, \bibinfo {author} {\bibfnamefont {C.}~\bibnamefont {Maier}},
  \bibinfo {author} {\bibfnamefont {J.}~\bibnamefont {Romero}}, \bibinfo
  {author} {\bibfnamefont {J.}~\bibnamefont {McClean}}, \bibinfo {author}
  {\bibfnamefont {T.}~\bibnamefont {Monz}}, \bibinfo {author} {\bibfnamefont
  {H.}~\bibnamefont {Shen}}, \bibinfo {author} {\bibfnamefont {P.}~\bibnamefont
  {Jurcevic}}, \bibinfo {author} {\bibfnamefont {B.~P.}\ \bibnamefont
  {Lanyon}}, \bibinfo {author} {\bibfnamefont {P.}~\bibnamefont {Love}},
  \bibinfo {author} {\bibfnamefont {R.}~\bibnamefont {Babbush}}, \bibinfo
  {author} {\bibfnamefont {A.}~\bibnamefont {Aspuru-Guzik}}, \bibinfo {author}
  {\bibfnamefont {R.}~\bibnamefont {Blatt}}, \ and\ \bibinfo {author}
  {\bibfnamefont {C.~F.}\ \bibnamefont {Roos}},\ }\href {\doibase
  10.1103/PhysRevX.8.031022} {\bibfield  {journal} {\bibinfo  {journal} {Phys.
  Rev. X}\ }\textbf {\bibinfo {volume} {8}},\ \bibinfo {pages} {031022}
  (\bibinfo {year} {2018})}\BibitemShut {NoStop}%
\bibitem [{\citenamefont {Ryabinkin}\ \emph {et~al.}(2018)\citenamefont
  {Ryabinkin}, \citenamefont {Yen}, \citenamefont {Genin},\ and\ \citenamefont
  {Izmaylov}}]{Ryabinkin:2018/qcc}%
  \BibitemOpen
  \bibfield  {author} {\bibinfo {author} {\bibfnamefont {I.~G.}\ \bibnamefont
  {Ryabinkin}}, \bibinfo {author} {\bibfnamefont {T.-C.}\ \bibnamefont {Yen}},
  \bibinfo {author} {\bibfnamefont {S.~N.}\ \bibnamefont {Genin}}, \ and\
  \bibinfo {author} {\bibfnamefont {A.~F.}\ \bibnamefont {Izmaylov}},\
  }\href@noop {} {\bibfield  {journal} {\bibinfo  {journal} {J. Chem. Theory
  Comput.}\ }\textbf {\bibinfo {volume} {14}},\ \bibinfo {pages} {6317}
  (\bibinfo {year} {2018})}\BibitemShut {NoStop}%
\bibitem [{\citenamefont {Jena}\ \emph {et~al.}(2019)\citenamefont {Jena},
  \citenamefont {Genin},\ and\ \citenamefont {Mosca}}]{MoscaA}%
  \BibitemOpen
  \bibfield  {author} {\bibinfo {author} {\bibfnamefont {A.}~\bibnamefont
  {Jena}}, \bibinfo {author} {\bibfnamefont {S.}~\bibnamefont {Genin}}, \ and\
  \bibinfo {author} {\bibfnamefont {M.}~\bibnamefont {Mosca}},\ }\href@noop {}
  {\bibfield  {journal} {\bibinfo  {journal} {arXiv.org}\ ,\ \bibinfo {pages}
  {arXiv:1907.07859}} (\bibinfo {year} {2019})}\BibitemShut {NoStop}%
\bibitem [{\citenamefont {Yen}\ \emph {et~al.}(2019)\citenamefont {Yen},
  \citenamefont {Verteletskyi},\ and\ \citenamefont {Izmaylov}}]{ThomsonA}%
  \BibitemOpen
  \bibfield  {author} {\bibinfo {author} {\bibfnamefont {T.-C.}\ \bibnamefont
  {Yen}}, \bibinfo {author} {\bibfnamefont {V.}~\bibnamefont {Verteletskyi}}, \
  and\ \bibinfo {author} {\bibfnamefont {A.~F.}\ \bibnamefont {Izmaylov}},\
  }\href@noop {} {\bibfield  {journal} {\bibinfo  {journal} {arXiv.org}\ ,\
  \bibinfo {pages} {arXiv:1907.09386}} (\bibinfo {year} {2019})}\BibitemShut
  {NoStop}%
\bibitem [{\citenamefont {Izmaylov}\ \emph {et~al.}(2020)\citenamefont
  {Izmaylov}, \citenamefont {Yen}, \citenamefont {Lang},\ and\ \citenamefont
  {Verteletskyi}}]{Izmaylov2019UP}%
  \BibitemOpen
  \bibfield  {author} {\bibinfo {author} {\bibfnamefont {A.~F.}\ \bibnamefont
  {Izmaylov}}, \bibinfo {author} {\bibfnamefont {T.-C.}\ \bibnamefont {Yen}},
  \bibinfo {author} {\bibfnamefont {R.~A.}\ \bibnamefont {Lang}}, \ and\
  \bibinfo {author} {\bibfnamefont {V.}~\bibnamefont {Verteletskyi}},\
  }\href@noop {} {\bibfield  {journal} {\bibinfo  {journal} {J. Chem. Theory
  Comput.}\ }\textbf {\bibinfo {volume} {16}},\ \bibinfo {pages} {1055}
  (\bibinfo {year} {2020})}\BibitemShut {NoStop}%
\bibitem [{\citenamefont {Huggins}\ \emph {et~al.}(2019)\citenamefont
  {Huggins}, \citenamefont {McClean}, \citenamefont {Rubin}, \citenamefont
  {Jiang}, \citenamefont {Wiebe}, \citenamefont {Whaley},\ and\ \citenamefont
  {Babbush}}]{BabbushA}%
  \BibitemOpen
  \bibfield  {author} {\bibinfo {author} {\bibfnamefont {W.~J.}\ \bibnamefont
  {Huggins}}, \bibinfo {author} {\bibfnamefont {J.}~\bibnamefont {McClean}},
  \bibinfo {author} {\bibfnamefont {N.}~\bibnamefont {Rubin}}, \bibinfo
  {author} {\bibfnamefont {Z.}~\bibnamefont {Jiang}}, \bibinfo {author}
  {\bibfnamefont {N.}~\bibnamefont {Wiebe}}, \bibinfo {author} {\bibfnamefont
  {K.~B.}\ \bibnamefont {Whaley}}, \ and\ \bibinfo {author} {\bibfnamefont
  {R.}~\bibnamefont {Babbush}},\ }\href@noop {} {\bibfield  {journal} {\bibinfo
   {journal} {arXiv.org}\ ,\ \bibinfo {pages} {arXiv:1907.13117}} (\bibinfo
  {year} {2019})}\BibitemShut {NoStop}%
\bibitem [{\citenamefont {Gokhale}\ \emph {et~al.}(2019)\citenamefont
  {Gokhale}, \citenamefont {Angiuli}, \citenamefont {Ding}, \citenamefont
  {Gui}, \citenamefont {Tomesh}, \citenamefont {Suchara}, \citenamefont
  {Martonosi},\ and\ \citenamefont {Chong}}]{ChicagoA}%
  \BibitemOpen
  \bibfield  {author} {\bibinfo {author} {\bibfnamefont {P.}~\bibnamefont
  {Gokhale}}, \bibinfo {author} {\bibfnamefont {O.}~\bibnamefont {Angiuli}},
  \bibinfo {author} {\bibfnamefont {Y.}~\bibnamefont {Ding}}, \bibinfo {author}
  {\bibfnamefont {K.}~\bibnamefont {Gui}}, \bibinfo {author} {\bibfnamefont
  {T.}~\bibnamefont {Tomesh}}, \bibinfo {author} {\bibfnamefont
  {M.}~\bibnamefont {Suchara}}, \bibinfo {author} {\bibfnamefont
  {M.}~\bibnamefont {Martonosi}}, \ and\ \bibinfo {author} {\bibfnamefont
  {F.~T.}\ \bibnamefont {Chong}},\ }\href@noop {} {\bibfield  {journal}
  {\bibinfo  {journal} {arXiv.org}\ ,\ \bibinfo {pages} {arXiv:1907.13623}}
  (\bibinfo {year} {2019})}\BibitemShut {NoStop}%
\end{thebibliography}
%

\section*{Supplementary Material: Hamiltonian details}


\paragraph*{H$_2$ molecule:}
One- and two-electron integrals in the canonical \gls{RHF} molecular
orbitals basis for $R$(H-H)=1.5 \AA, 
were used in the \gls{BK} transformation to produce the corresponding qubit Hamiltonian.
Spin-orbitals were alternating in the order $\alpha$, $\beta$,
$\alpha$, .... 

\paragraph*{LiH molecule:}
We will consider the LiH molecule at $R{\rm (Li-H)}=3.2$ \AA, it 
has a 6-qubit Hamiltonian containing 118 Pauli words. 
It was generated using the parity fermion-to-qubit transformation.
Spin-orbitals were arranged as ``first all alpha then all beta'' in
the fermionic form; since there are 3 active molecular orbitals in the
problem, this leads to 6-qubit Hamiltonian. 

This qubit Hamiltonian has 3$^{rd}$ and 6$^{th}$
stationary qubits, which allows one to replace the corresponding
$\hat z$ operators by their eigenvalues, $\pm 1$, thus defining the
different ``sectors'' of the original Hamiltonian. Each of these
sectors is characterized by its own 4-qubit effective Hamiltonian. The
ground state lies in the $z_{3} = -1$, $z_{6} = 1$ sector; 
the corresponding 4-qubit effective Hamiltonian ($\hat H_{\rm LiH}$) has
100 Pauli terms.  

\paragraph*{\ce{H2O} molecule:}
6- and 26-qubit Hamiltonians were generated for this system in the 6-31G basis, and the
14-qubit Hamiltonian was generated using the STO-3G basis. 
The geometry for all Hamiltonians was chosen to be $R{\rm (O-H)}=0.75$ \AA~ and $\angle HOH = 107.6^{\circ}$
The 14- and 26-qubit Hamiltonians were obtained in OpenFermion using both JW and BK transformations without any modifications, while for the 6-qubit Hamiltonian we used several qubit reduction techniques detailed below. 

Complete active space (4,\,4) electronic Hamiltonians at different \ce{O-H}
  distances were converted to the qubit form using the \gls{BK}
  transformation grouping spin-orbitals as ``first all alpha than all
  beta''. The resulting 8-qubit Hamiltonians contained 185 Pauli terms.
  $4^\text{th}$ and $8^\text{th}$ qubits were found to be stationary;
  the ground state solution is located in the $z_{3} = 1$, $z_{7} = 1$
  subspace. By integrating out $z_{3}$ and $z_{7}$,  a 6-qubit reduced 
  Hamiltonian with 165 terms was derived.  

\paragraph*{\ce{N2} molecule:} The BK and JW transformations of the electronic Hamiltonian in the 6-31G and STO-3G
bases produced 36- and 20-qubit Hamiltonians by OpenFermion, $R{\rm (N-N)}=1.1$ \AA. 

\paragraph*{\ce{BeH2} molecule:} The BK and JW transformations of the fermionic Hamiltonians in the 6-31G and STO-3G bases produced 26- and 14-qubit Hamiltonians by OpenFermion, $R{\rm (Be-H)} = 1.4 $ \AA, collinear geometry. 

\paragraph*{\ce{NH3} molecule:} The BK and JW transformations of the fermionic Hamiltonian in the 6-31G and STO-3G bases produced 30- and 16-qubit Hamiltonians by OpenFermion, $\angle HNH = 107^{\circ}$ and  $R{\rm (N-H)} = 1.0 $ \AA. 

\end{document}